\documentclass{iau307}
\usepackage{graphicx}
\usepackage{natbib}
\usepackage{url}
\usepackage{dtklogos}
\bibpunct{(}{)}{;}{a}{}{,}

\title[Massive star archeology in GCs] 
{Massive star archeology in globular clusters}

\author[W. Chantereau, C. Charbonnel \& G. Meynet]   
{W. Chantereau$^1$, C. Charbonnel$^{1,2}$
 \and G. Meynet$^1$}

\affiliation{$^1$Departement of Astronomy, University of Geneva, 1290 Versoix, Switzerland \\ email: {\tt william.chantereau@unige.ch} \\[\affilskip]
$^2$IRAP, CNRS UMR 5277, Universit\'e de Toulouse, 31400 Toulouse, France}

\pubyear{2014}
\volume{307} 
\pagerange{}
\jname{New windows on massive stars: asteroseismology, interferometry, and spectropolarimetry}
\editors{G. Meynet, C. Georgy, J.H. Groh \& Ph. Stee, eds.}

\begin{document}

\maketitle

\begin{abstract}
Globular clusters are among the oldest structures in the Universe and they host today low-mass stars and no gas. However, there has been a time when they formed as gaseous objects hosting a large number of short-lived, massive stars. 
Many details on this early epoch have been depicted recently through unprecedented dissection of low-mass globular cluster stars via spectroscopy and photometry.
In particular, multiple populations have been identified, which bear the nucleosynthetic fingerprints of the massive hot stars long disappeared.
Here we discuss how massive star archeology can been done through the lens of these multiple populations.

\end{abstract}

\firstsection 
\section{Introduction and the FRMS scenario}

Observations of Globular Clusters (GCs) show strong star-to-star variations of light elements (C, N, O, Na, Mg, Al) while iron and heavier elements stay constant (except in the most massive cases like $\omega$Cen, M22, NGC 1851, NGC 2419 and M54). They also reveal multiple structures (MS, SGB \& RGB) or even extended HBs in the GCs CMDs. This suggests that these stars clusters are composed of at least two populations of stars: the first one has a chemical composition similar to that of halo stars, whereas the second one is characterized by a very peculiar composition.
According to the FRMS scenario \citep{Decressin07,Prantzos06,Krause13}, fast rotating massive stars (25-120 M$_{\odot}$) have polluted the intracluster medium with their H-burning products ejecta from which a second generation of stars was born, bearing the fingerprint of the massive stars. In particular, their He content is expected to cover a large range between 0.248 (initial) and 0.8 (for the most O-depleted and Na-enriched ones).  

\section{The second generation\label{SecTwo}}

\begin{figure}[t]
\begin{center}
\includegraphics[width=\textwidth]{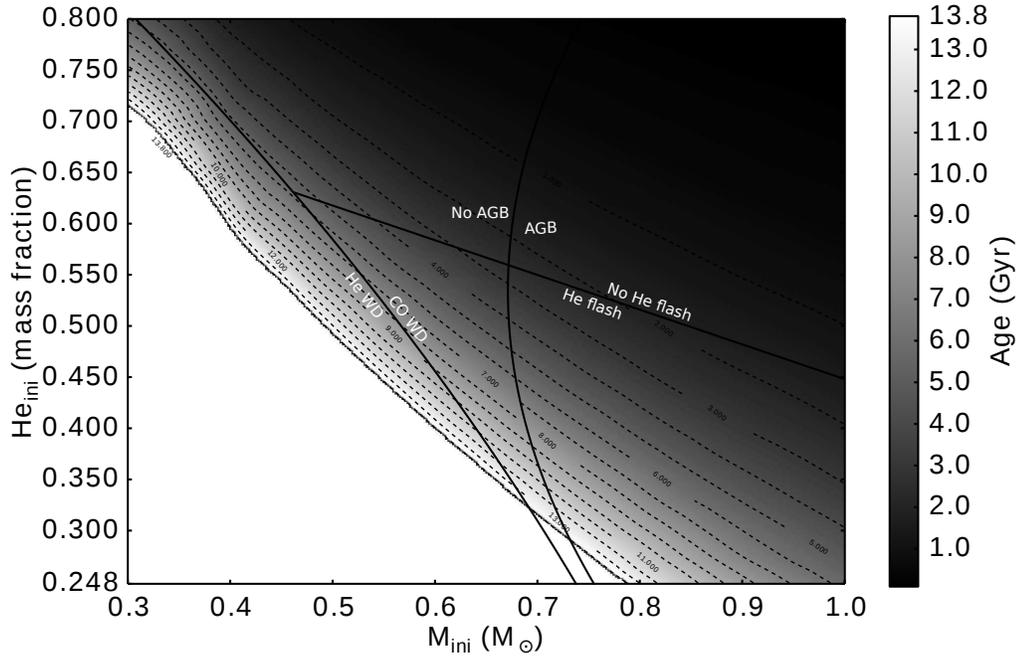} 
\caption{Lifetime and fate of stars versus initial mass and helium content at the
metallicity of NGC 6752. The colour code represents the age of the star at the turnoff in Gyr, the blank part represents a time $>$ 13.8 Gyr.}
\label{fig1}
\end{center}
\end{figure}

We compute a grid of standard stellar models (i.e. with no rotation nor atomic diffusion) for the second generation low-mass stars at low metallicity, the assumed initial chemical composition reflects directly the ejecta pollution from the FRMS (i.e. a higher initial helium content but also the anticorrelation CN, ONa, and MgAl), diluted at various degrees with the intracluster gas. This peculiar initial composition influences greatly the stellar evolution, making these stars hotter and brighter for each evolution phase.
Therefore these stars evolve faster; e.g. the age at the turnoff for a 0.8 M$_{\odot}$ with an initial He content $Y = 0.248$ is $\sim$13 Gyr when it is only $\sim$161 Myr for a 0.8 M$_{\odot}$ with $Y = 0.8$. Additionally the coupled effect of the initial mass and initial helium brings a wide variety of behaviors not usually expected for this range of mass, e.g. the presence of helium white dwarfs at a time lower than the Hubble one, and exacerbated late hot flashers/AGB-manqu\'e behaviors, see Fig. \ref{fig1}). 

\section{Direct application\label{SecThree}}

These second generation stars greatly impact the morphology of the current GCs because of their very peculiar composition but also because they outnumber the first generation of stars in GCs \citep{Prantzos06,Carretta10}. For instance \cite{Campbell13} observed that all the Na-rich stars (associated to the second generation, $\sim$70\%) fail to reach the AGB phase in NGC 6752. A straightforward explanation has been presented by \cite{Charbonnel13} who showed that all the Na-rich stars ([Na/Fe]$>0.4$ dex) within the framework of the FRMS scenario do not ascend the AGB, this is the so-called AGB-manqu\'e phenomenon.

\bibliographystyle{iau307}
\bibliography{Chantereau}

\end{document}